\documentclass{article}
\usepackage{spconf,amsmath,graphicx,url}

\usepackage[utf8]{inputenc}
\usepackage[english,russian]{babel}
\usepackage[LFE,LAE]{fontenc}

\usepackage{arabtex}
\usepackage{utf8}
\usepackage{footnote}
\makesavenoteenv{tabular}
\makesavenoteenv{table}

\setcode{utf8}

\title{VoxLingua107: a Dataset for Spoken Language Recognition}
\name{Jörgen Valk, Tanel Alumäe}
\address{Tallinn University of Technology, Estonia}

\begin{document}
\selectlanguage{english}
\ninept
\maketitle
\begin{abstract}
This paper investigates the use of automatically collected web audio data
for the task of spoken language recognition.
We generate semi-random search phrases from language-specific Wikipedia data that are 
then used to retrieve videos from YouTube for 107 languages.
Speech activity detection and
speaker diarization are used to extract segments from the videos that contain
speech. Post-filtering is used to remove segments from the database that are likely 
not in the given language, increasing the proportion of correctly labeled segments to 98\%, based on crowd-sourced verification. The size of the resulting training set (VoxLingua107) is 6628 hours (62 hours per language on the average) and it is accompanied by an evaluation set of 1609 verified utterances.
We use the data to build language recognition models for several spoken language identification
tasks. Experiments show that using the automatically retrieved training data gives competitive
results to using hand-labeled proprietary datasets. The dataset is publicly available\footnote{\url{http://bark.phon.ioc.ee/voxlingua107/}}.
\end{abstract}


\section{Introduction}

Spoken language recognition (SLR) is the task of automatically classifying an utterance based on the spoken language. SLR is used as a pre-processing step in several applications, such as automatic call routing, multilingual spoken translation and human-machine communication systems, multilingual speech transcription systems and spoken document retrieval. SLR is also often used in the area of intelligence and security.

In the past 20 years, development of SLR technology has been largely fostered through NIST Language Recognition Evaluations (LREs). As a result, the most popular benchmarks for evaluating new SLR models and methods are NIST LRE evaluation datasets \cite{sadjadi20182017}. The NIST LRE evaluations datasets contain mostly narrow-band conversational telephone speech. In order to build competitive systems for NIST LREs, large amounts of conversational telephone data from the particular languages are used. Such datasets are typically distributed by the Linguistic Data Consortium (LDC) and cost thousands of dollars. For example, the standard Kaldi \cite{kaldi} recipe for LRE07\footnote{\url{https://github.com/kaldi-asr/kaldi/blob/master/egs/lre07/v2/run.sh}} relies on 18 LDC SLR datasets that cost \$15400 to LDC non-members\footnote{The datasets were provided for free to the LRE07  particpants.}. This makes it difficult for new research groups to  enter the academic field of SLR. Furthermore, the NIST LRE evaluations focus mostly on telephone speech. There are not many resources for objectively evaluating SLR technology for other kinds of speech, such as conversational speech ``from the wild'', i.e., from various internet resources. Another problem lies in language coverage: NIST LREs have put a lot of focus on challenging classification tasks (e.g., fine-grained distinction of Arabic and Spanish dialects), but achieving a very large language coverage is not its main goal.  

The aim of this work is to investigate, whether automatically scraped and labelled speech data from the web can be used for building SLR systems. Our goal is to target various wide-band acoustic conditions and provide a large language coverage. We extract audio data from YouTube videos that are retrieved using random language-specific search phrases. If the language of the video title and description matches with the language of the search phrase, the audio in the video is likely to be in that particular language. This allows to collect large amounts of somewhat noisy data relatively cheaply.

There are several previous works that use an approach that is similar to ours. The KALAKA-3 database \cite{rodriguez2016kalaka} contains varying amounts of data from YouTube for several European languages. The YouTube data was retrieved semi-automatically: for each language, a list of search words was created based on the language dictionary, and YouTube was searched using single words from this list. Geographical location tags associated to some of the videos were also used to increase the chances of finding speech in a particular language. All the retrieved videos were hand-validated. Another dataset that uses YouTube audio is ADI17 \cite{shon2020adi17}, built for the MGB-5 Challenge \cite{ali2019mgb} Arabic dialect identification task. The dataset contains 3000 hours of speech from 17 Arabic dialects and is collected by manually compiling lists of popular YouTube channels in specific countries, and then retrieving videos from the corresponding channels. Using large amounts of YouTube audio is common in speaker recognition. The Speakers in the Wild (SITW) database \cite{mclaren2016speakers} contains speech samples of nearly 300 well-known public figures from open-source media. The research group behind SITW also describe a similar in-house dataset for language recognition which contains 2180 audio files sourced from open-source videos and represents 29 languages \cite{mclaren2018approaches}, but it has never been publicly released. The VoxCeleb 1\&2 \cite{nagrani2017voxceleb, Chung18b} corpora are popular free datasets for training speaker recognition models, containing automatically retrieved data from YouTube. The datasets have been compiled by using a phrase ``\textit{$<$celebrity name$>$ interview}'' as a YouTube search phrase and then extracting the segments from the retrieved videos where the corresponding celebrity is speaking, using face identification and lip synchronization detection.

\section{Scraping for spoken language data}

This section covers the process and tools that were used for collecting the speech data that
was used to build the SLR models.

\subsection{Method}

\begin{figure}[t]
  \centering
  \includegraphics[width=\linewidth]{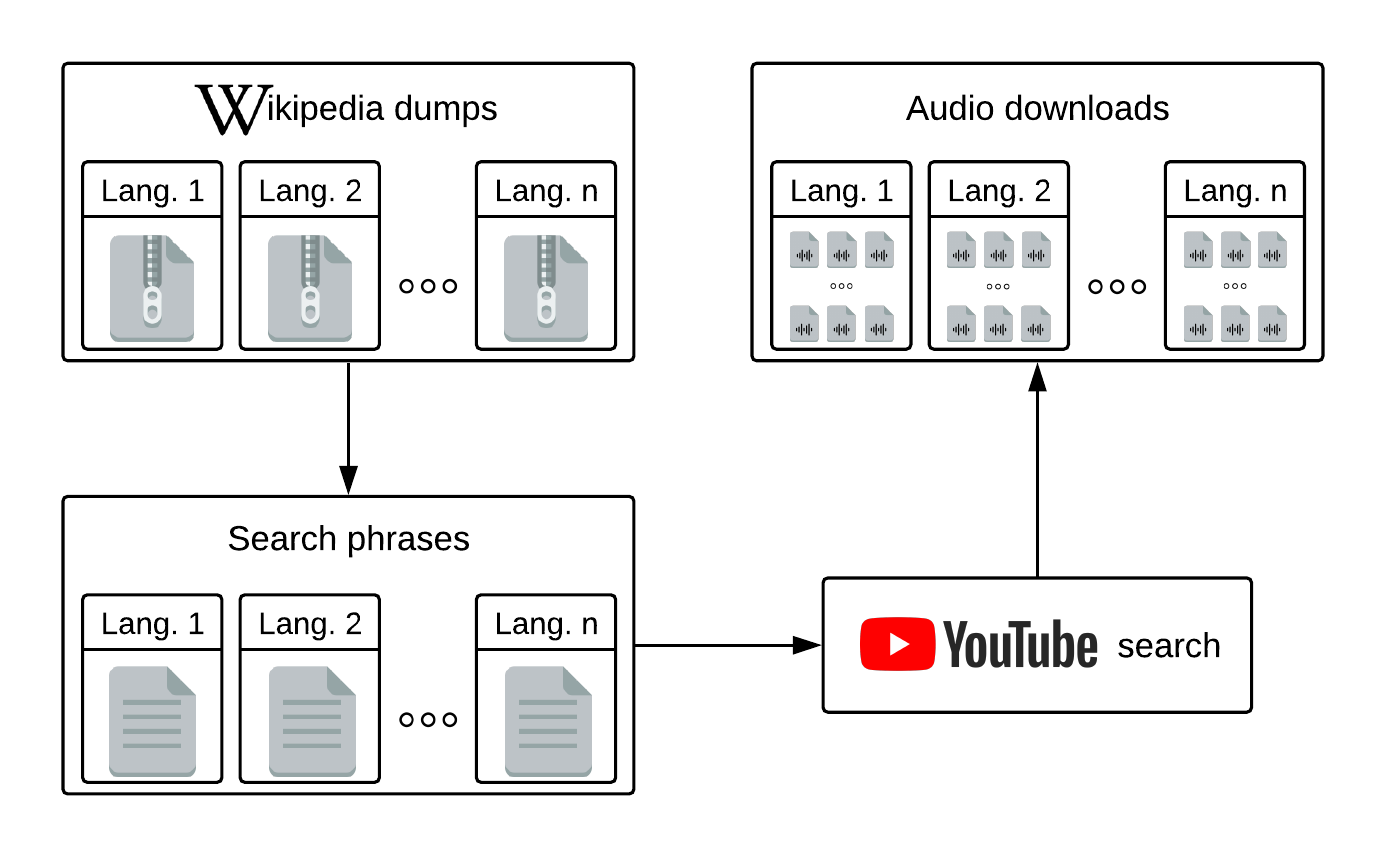}
  \caption{High level overview of the data collection process.}
  \label{fig:process}
\end{figure}

In order to build SLR models, we need a large corpus of speech utterances that is labeled according to the spoken language. In this work, we use  two different
data sources to automatically collect such data: Wikipedia and YouTube. We rely on Wikipedia to extract
random search phrases for each language, and on YouTube for retrieving videos whose title or metadata matches 
the search phrase and which is thus likely in the particular language. 
Such automatic data collection
process will undeniably not be 100\% accurate and there will be false positive video results
that have to be dealt with.

The data collection process can be divided into multiple steps. A diagram describing the process and steps is shown in Figure \ref{fig:process}. 

\subsubsection{Generation of search phrases}

We use Wikipedia dumps\footnote{\url{https://dumps.wikimedia.org}} for generating random language-specific search phrases. 
The number and quality of articles in different language's Wikipedia has a lot of variety.
For some smaller or less active language communities there may be less than 1\,000 articles. 
After some initial experiments it was decided that only languages that have more than 10\,000 articles are appropriate for this method.
Out of the more than 300 languages available in Wikipedia, 149 have more than 10\,000
articles. However, in some cases the limit of 10\,000 may not be enough because in some languages,
a large amount of articles are automatically generated, translated, or the contents are mostly
about certain kind of entities (such as geographical locations) which are not suitable for random search phrase generation.
We filtered out articles shorter than 3\,000 characters and the ones containing just a title. This improved the results in the next phrase
generation step. 

There are several different approaches that could be used to generate the search phrases from Wikipedia data.
The method that was used in this work is based on Term
Frequency Inverse Document Frequency (TF-IDF). TF-IDF is a well known method that
helps to evaluate how important a word or phrase in a document is. 
We applied TF-IDF on each language's Wikipedia dataset and compiled  a long list
of the ``most important'' phrases for each of the languages.

After doing some initial experiments with the collected datasets in different languages,
it was concluded that a good and universal phrase length to use in the YouTube search
engine is three words. Using shorter phrases resulted in vague results that also had
many false positives. Longer phrases, on the other hand, decreased the amount of search matches too much.

However, even with three word search phrases, there were still many false positives.
Many of the false positives were caused by phrases that were in other language than expected or that contained too many
numbers, stop-words, etc. To alleviate the main problem of the phrases being in the
wrong language, we used a 
text-based language identification model on the generated phrases to filter out
all phrases whose language was unknown or did not match the expected language.
We used the Polyglot\footnote{\url{https://polyglot.readthedocs.io}} Python package 
for language identification which supports 165 languages. 
This step removed a relatively large portion of the generated phrases and resulted in  the final set of
phrases that deemed usable for the next step. A randomly picked sample of the generated
phrases for eight languages can be seen in Table \ref{tab:example-phrases}.

\begin{table}[tbh]
\caption{Sample random search phrases.}
\label{tab:example-phrases}
\centering
\begin{tabular}{lr}
Language & Random search phrase \\
\hline
English  & \textit{the northern territory} \\
Estonian & \textit{ameerika ühendriikide relvajõud} \\
Finnish  & \textit{hiileen liittynyt hydroksyyliryhmä} \\
German   & \textit{abgesetzten enameloliden zahnkappen} \\
Latvian  & \textit{starptautiskajā šaha turnīrā} \\
Russian  & \textit{\foreignlanguage{russian}{совета рабочих депутатов}} \\
Spanish  & \textit{administración del estado} \\
Urdu     & \textit{\<انیہ اور فرانس >} \\
\hline
\end{tabular}
\end{table}

\begin{figure}[t]
  \centering
  \includegraphics[width=\linewidth]{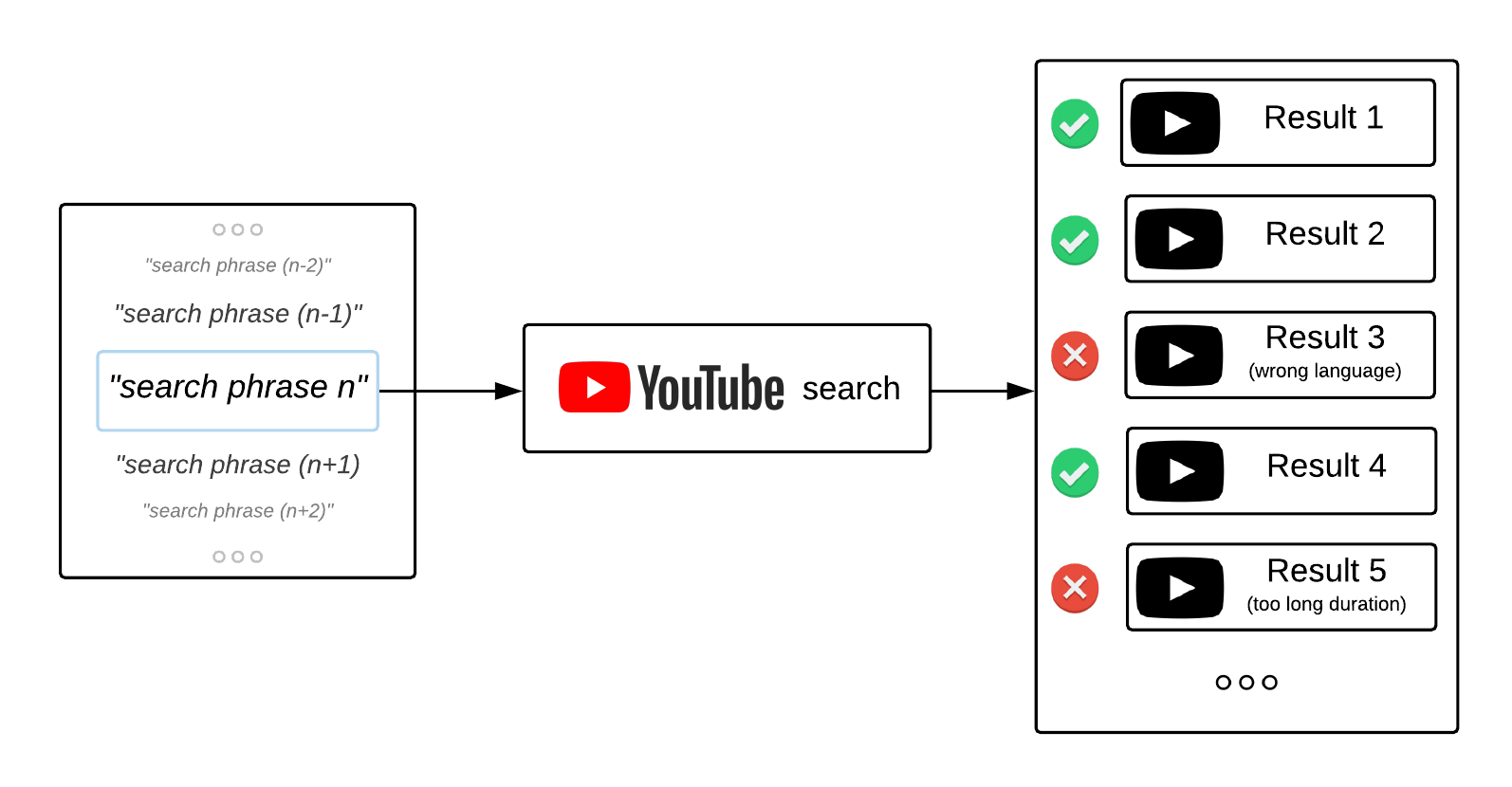}
  \caption{Overview of the process of retrieving and filtering of videos.}
  \label{fig:youtube}
\end{figure}

\subsubsection{Retrieving audio data}

The next step was using the generated phrases for collecting the videos from which the audio could be extracted.
An overview of the process is given in Figure \ref{fig:youtube}.
All of the previously generated phrases were used one by one to find matching
videos from YouTube. Even though the phrases were heavily processed and filtered, this still resulted
in many false positives, i.e., videos not in the expected language.
There are also cases where the video title is in the
expected language but the content is in another language.
To decrease the number of false positive
video results, we applied the text based language identification model to 
the video title, description (which usually reflects the video content
and language most closely) and other metadata, if available. All
results with unknown or not expected language were filtered out, similarly to the search
phrase generation process before. This decreased the amount of false positive results
noticeably. 

\subsubsection{Audio segmentation}

As the final step in the dataset generation process, audio files were split into shorter utterance-like segments and 
the resulting segments containing mostly noise, music or silence were removed. The LIUM SpkDiarization toolkit \cite{rouvier2013open} was applied for this. We imposed a limit of a minimium clip length of 2 seconds and a maximum clip length of 20 seconds. The majority of the resulting speech segments had a
duration between 4 to 10 seconds.

\subsection{Scraping results}

After the several filtering steps, we were able to download at least some videos for 107 languages. 
In total, audio was collected from close to 78\,000 videos, resulting in a dataset of 14\,044 hours of audio content (before segmentation and non-speech data removal). 

\begin{table*}[tbh]
\footnotesize
\caption{Total duration of cleaned data for each language in the training set, in hours (h).}
\label{tab:durs}
\centering
\begin{tabular}{lr|lr|lr|lr|lr|lr|lr}
\hline
Language    & h & Language  & h & Language    & h & Language      & h & Language   & h & Language  & h & Language   & h \\
\hline
Abkhazian   & 10    & Catalan   & 88    & German      & 39    & Kannada       & 46    & Marathi    & 85    & Shona     & 30    & Tibetan    & 101   \\
Afrikaans   & 108   & Cebuano   & 6     & Greek       & 66    & Kazakh        & 78    & Mongolian  & 71    & Sindhi    & 84    & Turkish    & 59    \\
Albanian    & 71    & C. Khmer  & 41    & Guarani     & 2     & Korean        & 77    & Nepali     & 72    & Sinhala   & 67    & Turkmen    & 85    \\
Amharic     & 81    & Chinese   & 44    & Gujarati    & 46    & Lao           & 42    & Norwegian  & 107   & Slovak    & 40    & Ukrainian  & 52    \\
Arabic      & 59    & Croatian  & 118   & Haitian     & 96    & Latin         & 67    & Nynorsk    & 57    & Slovenian & 121   & Urdu       & 42    \\
Armenian    & 69    & Czech     & 67    & Hausa       & 93    & Latvian       & 42    & Occitan    & 15    & Somali    & 103   & Uzbek      & 45    \\
Assamese    & 155   & Danish    & 28    & Hawaiian    & 12    & Lingala       & 90    & Panjabi    & 54    & Spanish   & 39    & Vietnamese & 64    \\
Azerbaijani & 58    & Dutch     & 40    & Hebrew      & 96    & Lithuanian    & 82    & Persian    & 56    & Sundanese & 64    & Waray      & 11    \\
Bashkir     & 58    & English   & 49    & Hindi       & 81    & Luxembourg. & 75    & Polish     & 80    & Swahili   & 64    & Welsh      & 76    \\
Basque      & 29    & Esperanto & 10    & Hungarian   & 73    & Macedonian    & 112   & Portuguese & 64    & Swedish   & 34    & Yiddish    & 46    \\
Belarusian  & 133   & Estonian  & 38    & Icelandic   & 92    & Malagasy      & 109   & Pushto     & 47    & Tagalog   & 93    & Yoruba     & 94    \\
Bengali     & 55    & Faroese   & 67    & Indonesian  & 40    & Malay         & 83    & Romanian   & 65    & Tajik     & 64    & \textbf{Total}      & \textbf{6628}  \\
Bosnian     & 105   & Finnish   & 33    & Interlingua & 3     & Malayalam     & 47    & Russian    & 73    & Tamil     & 51    & \textbf{Average}    & \textbf{62}    \\
Breton      & 44    & French    & 67    & Italian     & 51    & Maltese       & 66    & Sanskrit   & 15    & Tatar     & 103   &            &       \\
Bulgarian   & 50    & Galician  & 72    & Japanese    & 56    & Manx          & 4     & Scots      & 3     & Telugu    & 77    &            &       \\
Burmese     & 41    & Georgian  & 98    & Javanese    & 53    & Maori         & 34    & Serbian    & 50    & Thai      & 61    &            &      \\
\hline
\end{tabular}
\end{table*}



All of the collected videos have a maximum duration of one hour. The majority of the dataset consists of relatively
short clips between 1-10 minutes. This is positive, as it increases the variety
of speakers, acoustic conditions and topics.

After segmentation and speech/non-speech classification, the the number of utterances in the dataset is about 3.5 million.

\subsection{Validation using crowd-sourcing}

\begin{figure}[tb]
  \centering
  \includegraphics[width=\linewidth]{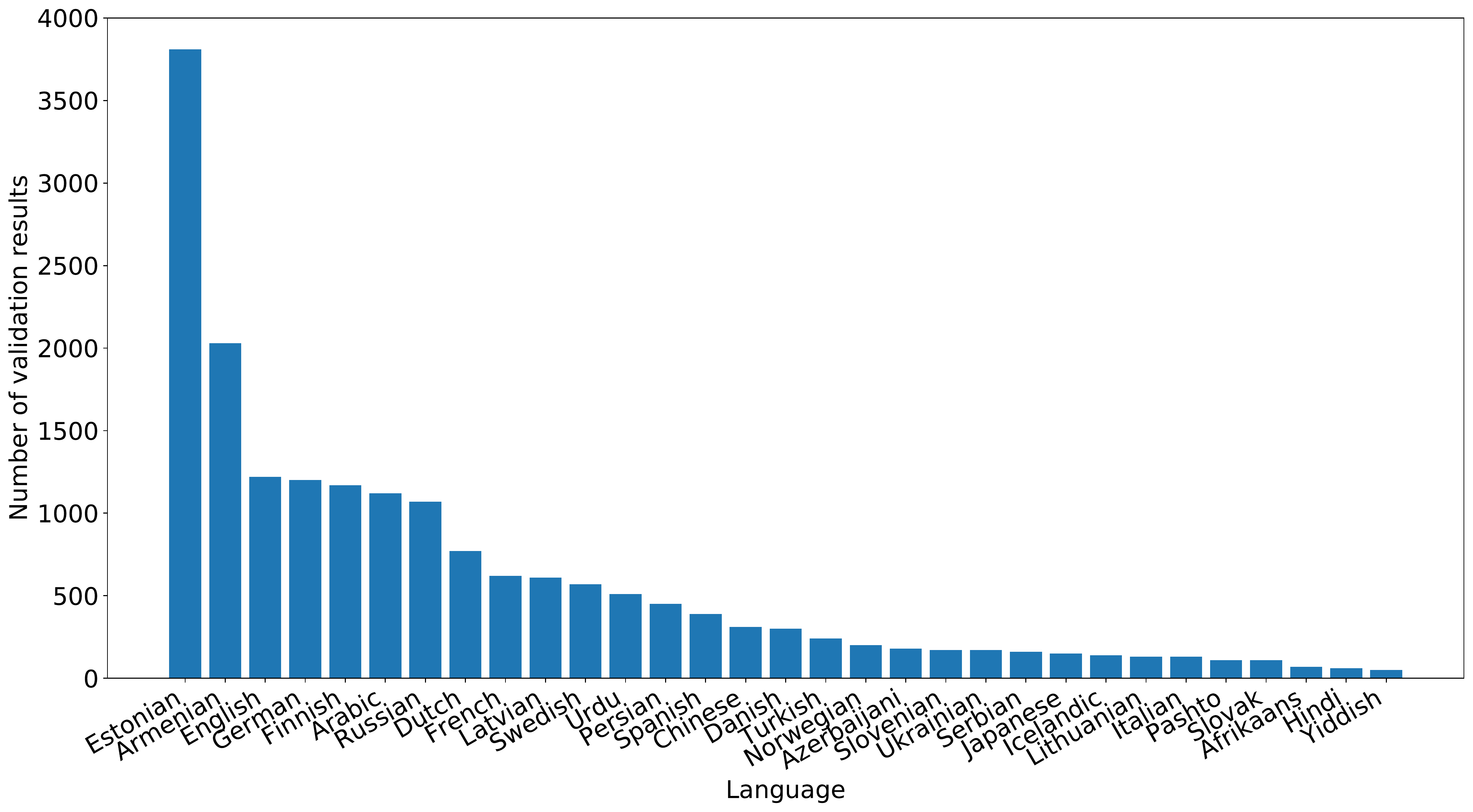}
  \caption{Number of human-provided labels for languages with more than 50 labels.}
  \label{fig:validation_per_language}
\end{figure}

Even after several filtering steps, the data collected using this method
still contains false positive results. False positives mostly occur when the language of the speech
in the video does not match with the language of the video title and description.

In order to assess the ratio of false positives in the dataset and to produce human-labelled
development data for subsequent SLR experiments, a crowd-sourcing based validation experiment was carried out. For this,
a custom web application was built that 
allowed to share the validation task to a wide group of people who are proficient in
different languages. The application first displays  a list of all the collected languages
and the user can pick a language to validate. The user's language proficiency on a scale of 1 to 5 is also asked. After selecting a language to validate, a random selection of ten audio clips is
presented to the user to work on. If possible, then some of the ten clips
are selected from the ones that have already been annotated once by another user. This would later allow us to analyze
inter-annotator agreement. For each presented clip, the user is asked to decide, whether the clip
(1) contains speech in the given language, (2) contains speech in some other language, 
(3) doesn't contain speech or (4) the user cannot provide a definite answer. 
Google account based authentication was added to the application in order to reduce the amount of
misbehaviour.

The main validation process was started in February 2020. We asked volunteers via social media
to contribute to our research. As of May 2020,  over 14\,000 unique audio segments have been 
labelled, with over 18\,000 labels in total. The amount of validation data per language varies a lot:
for some languages only a few audio clips have been validated. For  31 languages, more than 50 labels have been collected.
These 31 languages together with the corresponding counts are shown in Figure \ref{fig:validation_per_language}.


On the average, 85.3\% of the segments in our dataset contain speech in the expected language.
5.8\% utterances were in another language, 7.5\% contained non-speech and for 1.4\% utterances no definite answer was provided.
When not counting non-speech segments (which could be eliminated from the dataset using better speech/non-speech classification models) and answers where no definitive answer could be provided, the share of segments in the expected language increases to 93.6\%.
Figure \ref{fig:validation_per_language} shows the proportion of different validation labels per language. 

Average inter-annotator agreement of the validation results is 97.0\%, when ignoring the answers where no definitive answer was provided.

\begin{figure}[tb]
  \centering
  \includegraphics[width=\linewidth]{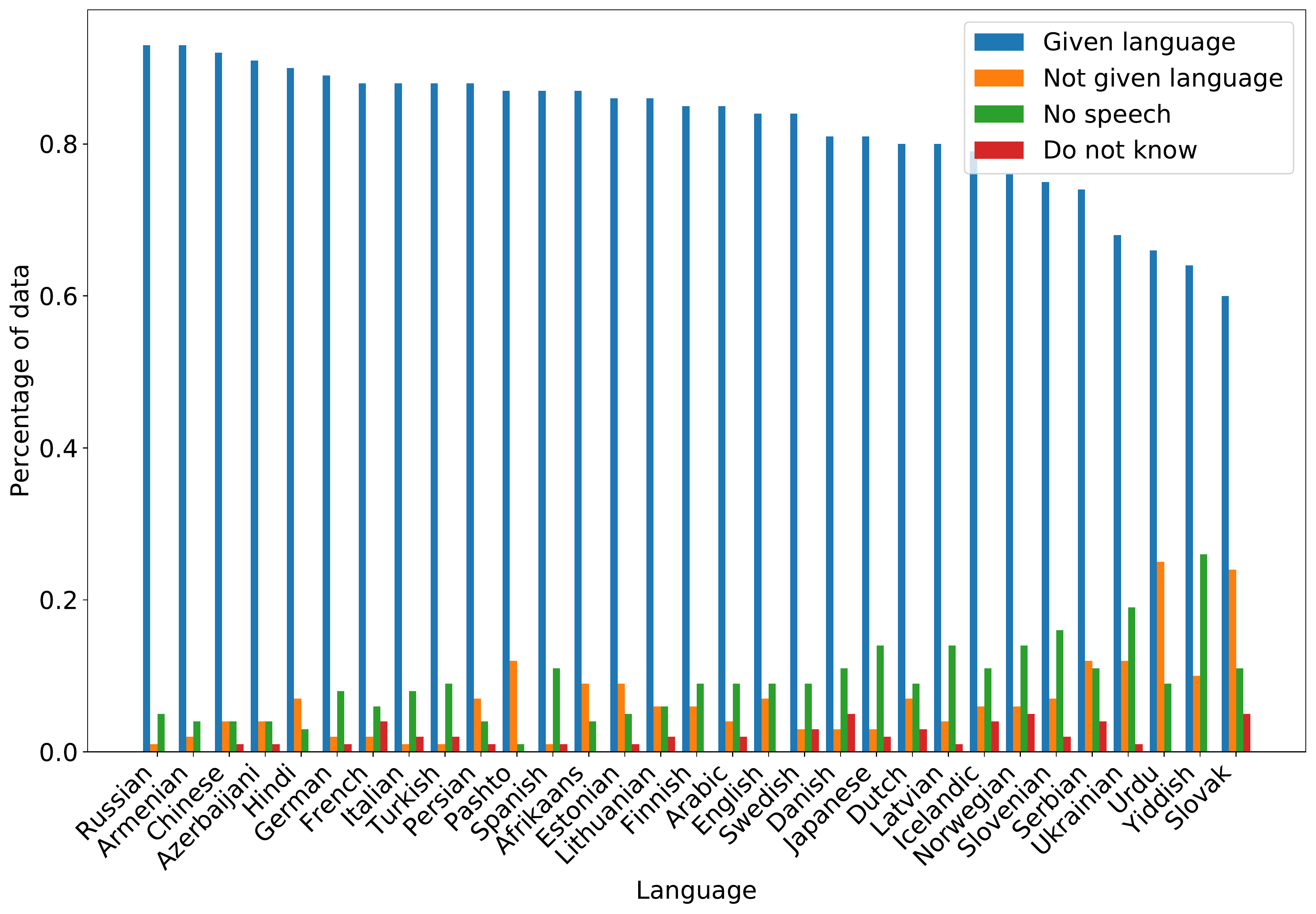}
  \caption{Distribution of crowd-sourced labels per language.}
  \label{fig:validation_answer_percentages}
\end{figure}


\subsection{Data-driven filtering}

As mentioned in the previous section, only an estimated 85\% of the segments in the automatically compiled dataset are in the expected language. To address this issue, we experimented with data-driven filtering. The filtering aims to remove segments that are probably not in the given language from the dataset, while keeping most of the valid segments.

We used a recently proposed technique called Robust Generative classifier (RoG) \cite{lee2019robust} for filtering. First, we trained an x-vector \cite{snyder2018x} language identification model on the original noisy data. The model was used to extract x-vector embeddings for all utterances. On top of the extracted features, we trained a generative classifier which utilizes the minimum covariance determinant (MCD) estimator to estimate its parameters, as proposed in \cite{lee2019robust}. Finally, we applied the resulting generative classifier on the full dataset and chose a posterior probability threshold which roughly equalizes the false positive rate (i.e., segments with false labels that are above the threshold) and the false negative rate (segments with correct labels falling below the threshold), using the small subset of the dataset with crowd-sourced labels. All segments which fell under the threshold were removed from dataset.

The filtering reduced the proportion of incorrectly labeled segments in the data from 15\% to 2\%, while retaining 90\% of the correctly labeled segments. In the next section we show that such filtering had a positive effect on the accuracy of the resulting language identification models.

We split the filtered dataset into official training and evaluation set. The evaluation set was created by selecting at most 100 utterances from each language that had been confirmed by at least two separate volunteers to be in the expected language. This amounts to 1609 segments from 33 languages (not all languages had validation data from two persons). To avoid data leakage, all utterances of the videos with at least one segment in the evaluation set were removed from the training set. Note that this does not remove the possibility of data leakage completely: the dataset might contain several videos from the same YouTube channel(s), with one or more of the videos in the training set and one in the evaluation set. This would allow the language identification model to use speaker recognition cues to perform well in the experiment. 

The total amount of data for each language in the cleaned training set is given in Table \ref{tab:durs}.


\section{Language identification experiments}

We conducted several experiments to assess the usefulness of the retrieved data for SLR. In all experiments, we use language embedding models derived from the x-vector paradigm \cite{snyder2018x,snyder2018spoken}, with several enhancements. During training, we apply on-the fly data augmentation using AugMix \cite{hendrycks2019augmix}, by randomly distorting the training data using a mix of reverberation and noise augmentation. For frame-level feature extraction, we use the Resnet34 \cite{cai2018exploring, he2016deep} architecture where the basic convolutional blocks with residual connections are replaced with squeeze-and-attention modules \cite{hu2018squeeze,zhou2019deep}.
For temporal pooling, we use multi-head attention, similarly to the method described in \cite{zhu2018self}. The models are implemented in PyTorch \cite{NEURIPS2019_9015} using a framework developed in our lab.

\subsection{In-domain data, 107-language classification}

For the SLR experiment with in-domain data, we trained two systems: one on the noisy dataset and one on the filtered dataset. For both systems, we followed the same procedure: first, a backend x-vector model was trained on the corresponding training dataset. The model was used to extract utterance embeddings for all training and evaluation data. The frontend LDA/PLDA based generative classifier was trained on the extracted x-vectors, using Kaldi's \cite{kaldi} implementation. 

The classification error rates of the two models is shown in Table \ref{tab:indomain}. We also calculated the error rates for short utterances (less than 5 seconds) and long utterances (5 to 20 seconds). Our dataset does not contain utterances longer than 20 seconds  since the preprocessing pipeline splits speech into to utterances with a maximum duration of 20 seconds.
As can be seen, the data-driven cleaning step does not improve classification performance on in-domain data. Table \ref{tab:indomain_err} shows most common classification errors encountered in the evaluation data. Most of the errors occur between closely related languages (in fact, Urdu and Hindi are mutually intelligible as spoken languages), although the substitution of English with Welsh is more likely caused by the bad quality of the retrieved Welsh data.

This system is available for experimentation at \url{https://bark.phon.ioc.ee/lid_demo}.

\begin{table}[tb]
\caption{Results on in-domain data. }
\label{tab:indomain}
\centering
\begin{tabular}{lccc}
\hline
Training data & \multicolumn{3}{c}{Error rate (\%)} \\
 & 0..5 sec  & 5..20 sec & Average \\
\hline
Noisy  & 12.3 & 6.1 & 7.1 \\
Cleaned & 13.4 & 6.6 & 7.6 \\
\hline
\end{tabular}
\end{table}

\begin{table}[tb]
\caption{Most common errors.}
\label{tab:indomain_err}
\centering
\begin{tabular}{c}
\hline
Urdu  $\xrightarrow{}$ Hindi \\
Spanish  $\xrightarrow{}$ Galician \\
Norwegian  $\xrightarrow{}$ Nynorsk \\
Dutch  $\xrightarrow{}$ Afrikaans  \\
English  $\xrightarrow{}$ Welsh  \\
Estonian   $\xrightarrow{}$ Finnish \\
\hline
\end{tabular}
\end{table}

\subsection{KALAKA-3}

The KALAKA-3 dataset \cite{rodriguez2016kalaka} consists of three partitions: training data, containing mostly broadcast speech for a 6 language subset (Basque, Catalan, English, Galician, Portugese and Spanish) \cite{rodriguezalbayzin},  development data and test data, containing YouTube audio for 21 European languages. There are four evaluation sets: Plenty-Closed (PC) and Plenty-Open (PO), handling closed and open set classification for the 6 listed languages, and Empty-Closed (EC) and Empty-Open  (EO), handling closed and open set classification for German, Greek, French and Italian. 

We experimented with several methods in this experiment. We trained two language embedding models, one on the noisy data and the other on the automatically filtered (cleaned) data. Both models were  trained on a 21 language subset of our  107 language dataset, covering all languages of the KALAKA-3 development and evaluation data (both evaluation and out-of-set languages).
For training the final logistic regression classifier, we experimented with two approaches.
In the first one, we trained the classifiers for each subtask on the x-vectors extracted from  our training dataset (noisy or cleaned). In the second approach, we used the KALAKA-3 development data (around 3 hours per in-set language) to train the final classifier. We didn't use the official KALAKA-3 training data in either of the approaches.

\begin{table}[tb]
\footnotesize
\caption{Results on KALAKA-3, in $F_{act}$/EER (lower is better).  }
\label{tab:kalaka3}
\centering
\begin{tabular}{p{1.0cm}p{0.8cm}cccc}
\hline
Embedding training data & LR train data & PC & PO & EC & EO \\
\hline
\multicolumn{2}{l}{Baseline \cite{rodriguez2016kalaka}} & .079/5.74 & .115/6.67 & .104/6.16 & .169/6.96 \\
Noisy & Noisy  &  .124/7.23 & .150/8.32  & .028/0.32 & .083/3.08 \\
Cleaned & Cleaned  &  .106/5.95 & .112/6.50  & .028/0.16 & .046/0.86 \\
Noisy &  Kalaka3 &  .055/4.36 & .083/5.95 & .033/0.32 & .059/3.68 \\
Cleaned  & Kalaka3  &  .041/3.51 &  .056/5.21 & .022/0.00 & .058/3.51 \\
\hline
\end{tabular}
\end{table}

Table \ref{tab:kalaka3} lists the evaluation results, using KALAKA-3 official performance metric $F_{act}$ and equal error rate (EER). As baseline, we use a fusion of various i-vector and phonotactic systems reported in \cite{rodriguez2016kalaka}. The results show that on the 4-language tasks (EC and EO), all of our approaches give relatively similar scores. On the 6 language task that mainly contains languages spoken in the Iberian peninsula (PC and PO), using the KALAKA-3 development data for training the final classifier gives notably better results. We suspect that this is because our data for the the four official languages in Spain is probably relatively noisy (e.g., our scraped Galician data likely contains a lot of Spanish, in spite of cleaning) and even very little amount of hand-labeled development data allows to build more accurate classifiers. We also see that also in this experiment, using filtered trainining data improves results. Note that our results are not directly comparable to the given baseline, since KALAKA-3 official rules allowed  using only the provided training data.

\subsection{LRE07}

The LRE07 dataset contains 14 languages that are used as detection
targets \cite{nist2007}. LRE07 dataset contains conversational telephone speech, which has
considerably different characteristics than the YouTube audio data collected in this
work.

We trained x-vector models on the 14 language subset of our scraped data that covers all the LRE07 languages, with audio resampled to 8 kHz. We used the resulting model to extract embeddings for our 
14 language training data subset  and
then built a logistic regression classifier, using centered and length-normalized embeddings. Since the audio in LRE07 evaluation data contains a lot of non-speech regions, we removed long regions of non-speech using a Kaldi speech activity detection model trained for the IARPA ASpIRE challenge by the JHU team \cite{peddinti2015jhu}, prior to extracting embeddings.
Table \ref{tab:lre07} shows the results of different systems.
All the compared systems are trained on NIST LRE training data, i.e., on large amounts of telephone speech.
It can be seen that our Resnet-based models, trained on out-of-domain data, achieve 
better average results than GMM, phonotactic and i-vector based models trained on in-domain data. Our models
perform particularly well on short utterances.
However, modern systems trained on in-domain data result in better performance. Our model
architecture is very similar to the CNN-SAP model \cite{cai2018exploring} which gives
around 16\% relative improvement compared to our system that is trained on automatically filtered data. Using the cleaned dataset shows benefits also in this experiment, except for the 30 second subset where the model trained on noisy data performs slightly better. 
\begin{table}[tb]
\caption{Results on LRE07  closed-set task, in $C_{avg}$ (lower is better).  }
\label{tab:lre07}
\centering
\begin{tabular}{lrrrr}
\hline
System & 3 sec & 10 sec & 30 sec & Average \\
\hline
\multicolumn{5}{l}{Trained on in-domain data (telephone speech)} \\
\hline
GMM-MMI \cite{torres2008mitll} & 17.28 & 5.90 & 2.10 & 8.42 \\
Fusion of models \cite{torres2008mitll} & 13.32 &  3.55 & 0.97 & 5.95 \\
Phonotactic \cite{gelly2016divide} & 18.59 & 6.28 & 1.34 & 8.73 \\
Fusion of models \cite{gelly2016divide} & 15.29 & 4.54 & 1.30 & 7.04 \\
Kaldi i-vector\footnote{\url{https://github.com/kaldi-asr/kaldi/blob/master/egs/lre07/v1/run.sh}}  & 26.04 &  11.93 &  4.52  & 14.17  \\
Kaldi i-vector DNN\footnote{\url{https://github.com/kaldi-asr/kaldi/blob/master/egs/lre07/v2/run.sh}}  & 19.67  & 7.84  & 3.31 & 10.27  \\
CNN-SAP \cite{cai2018exploring} & 8.59 & 2.49 & 1.09 & 4.06  \\
CNN-LDE \cite{cai2018exploring} & 8.25 & 2.61 & 1.13 & 4.00 \\
\hline
\multicolumn{5}{l}{Our systems trained on our dataset} \\
\hline
Resnet34 (noisy data)   & 10.58 & 3.33 & 1.72 & 5.21 \\
Resnet34 (cleaned data) & 9.39 & 3.14 & 1.90 & 4.81 \\
\hline
\end{tabular}
\end{table}

\section{Analysis of language embeddings}

\begin{figure*}[tbh]
  \centering
  \includegraphics[width=\linewidth]{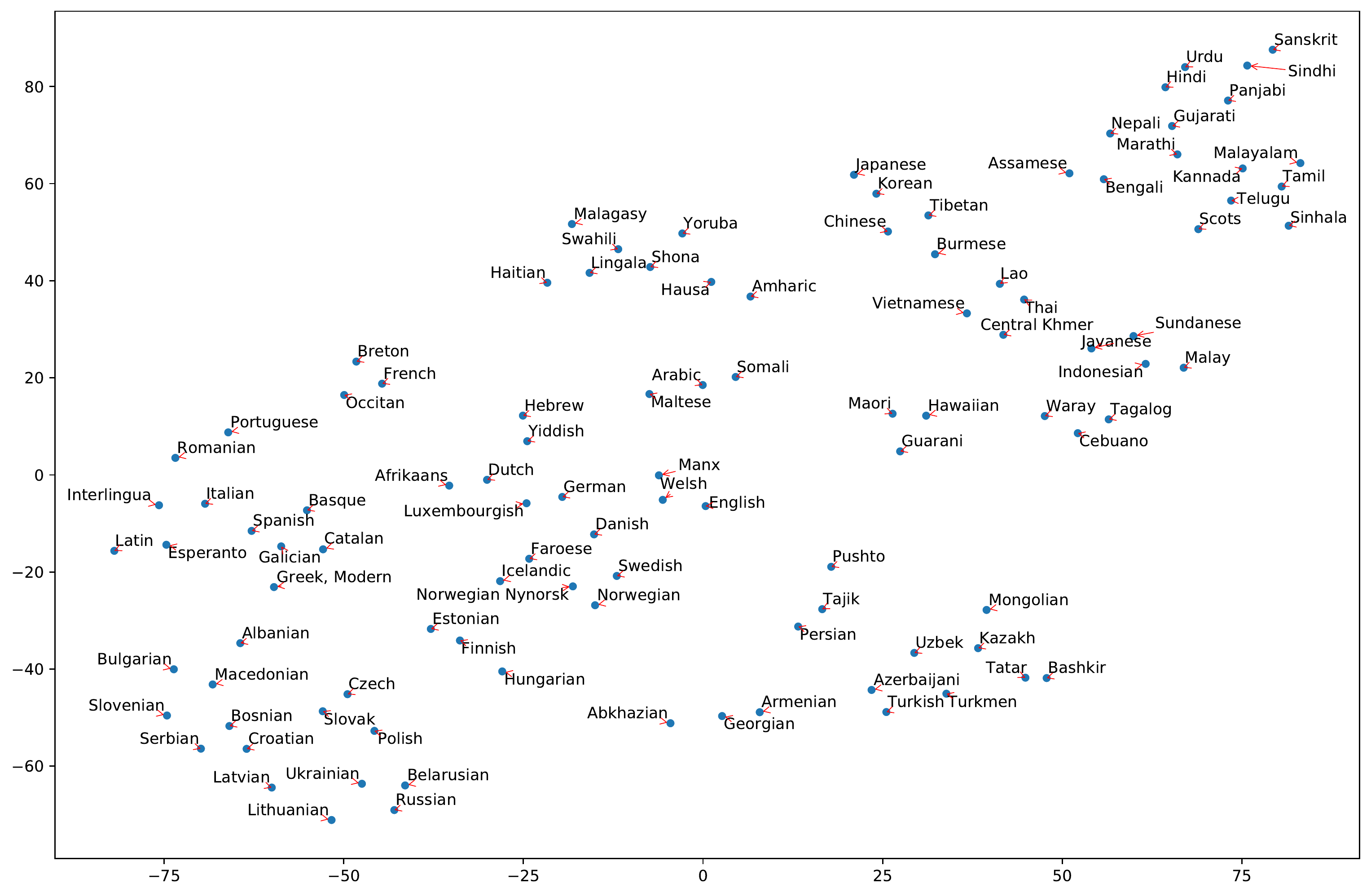}
  \caption{T-SNE plot of the language embeddings.}
  \label{fig:tsne}
\end{figure*}

After training an x-vector language recognition system, we extracted the embeddings of all utterances in the training data. LDA was used to reduce the dimensionality of the embeddings to 250. Language embeddings were then calculated by averaging language-specific utterances, transforming them to lower-dimensional space using LDA and applying length normalization.

Figure \ref{fig:tsne} shows the T-SNE plot of the resulting language embeddings.
It demonstrates that the learned language embeddings represent the structure of the language families remarkably well, although there are obvious deviations from the linguistic hierarchy, probably caused by cultural and geographical influences.

\section{Alternative approaches}


Using YouTube as the data source is not the only option for building a large-scale dataset for SLR.
The largest and most well-known free multilingual speech dataset is Mozilla Common Voice\footnote{\url{http://commonvoice.mozilla.org/}}. It contains prompted speech utterances donated by volunteers. Utterance metadata contains speaker identity, text of the prompt and language. Many utterances also include demographic metadata like age, sex, and accent. The dataset currently consists of 5671 validated hours in 54 languages. The main benefit of the this dataset over ours is that all the recordings in the validated corpus are confirmed by volunteers to really contain the given utterance. The main shortcoming of the corpus, with regard to using it for language recognition ``in the wild'', is that it contains only read speech. There is also a lot of variety in the amount available of data per language, and the language coverage is lower than in our data. Nevertheless, in the future we plan to experiment with combining Common Voice with our dataset.

Another practical alternative to our method is using podcasts as training data. 
Podcasts usually contain spontaneous speech which is beneficial for training SLR models.
Unfortunately, it seems that finding podcasts by language is not an easy task. There exists some domain-specific podcast listings by language (e.g., ``geek'' podcasts in 15 languages\footnote{\url{http://github.com/ayr-ton/awesome-geek-podcasts}}), but none of the large podcast directories allow browsing for podcasts by the language.

\section{Availability}

The dataset is publicly available at \url{http://bark.phon.ioc.ee/voxlingua107/}. Similarly to the VoxCeleb and ADI17 datasets, the VoxLingua107 dataset is distributed under the Creative Commons Attribution 4.0 International License. The copyright remains with the original owners of the video. 

While YouTube users own the copyright to their own videos, using the audio in the vidoes for training language identification models has very limited and transformative purpose and qualifies  thus as ``fair use'' of copyrighted materials. YouTube's terms of service forbid downloading, storing and distribution of videos. However, the aim of this rule is clearly to forbid unfair monetization of the content by third-party sites and applications. Our dataset contains the videos in segmented audio-only form that makes the monetization of the actual distributed content extremely difficult.

We also point out that the distribution of languages, accents, dialects, genders, races and societal factors in this dataset is not representative of the global population. Using this dataset for training and deploying models may thus introduce unintended biases.

\section{Conclusion}

We described a new speech dataset VoxLingua107 that is compiled from YouTube data. We showed that the dataset is suitable for training spoken language recognition models for classifying data ``from the wild''. Experiments with the KALAKA-3 dataset showed that the dataset can be used for building both a backend feature extractor and a frontend classifier, or only for training the backend, while the frontend is trained on small amounts of hand-labeled data.
Experiments on the NIST LRE07 evaluation data showed that using VoxLinugua107 data  results in a classifier that is not far in accuracy from a model trained on large amounts of in-domain data. 

Future work includes using the VoxLingua107 dataset together with in-domain training data, and augmenting input filterbank features with multilingual bottleneck features.

\bibliographystyle{IEEEbib}

\bibliography{mybib}

\end{document}